\documentclass[twocolumn,showpacs,preprintnumbers,amsmath,amssymb]{revtex4}

\usepackage{graphicx}
\usepackage{dcolumn}
\usepackage{bm}

\begin{document}


\title{An Alternative Analysis of the $R$ Measurements: Resonance Parameters of the Higher Vector States of Charmonium}

\author{Kamal K. Seth}
 \email{kseth@northwestern.edu}
\affiliation{%
Northwestern University, Evanston, Illinois 60208, USA
}%

\date{\today}

\begin{abstract}
The existing experimental data on total cross sections for hadron production in $e^{+}e^{-}$ annihilation in the resonance region $\sqrt{s}$ = 3.8 - 4.8 GeV, usually presented in terms of the parameter $R$, are critically examined. It is shown that the Crystal Ball and BES measurements are in excellent agreement, and their analysis leads to consistent resonance parameters for the three vector resonances above the $D\bar D$ threshold. The results for the widths are found to be considerably different from the presently adopted values, and have much smaller errors.
\end{abstract}

\maketitle

One of the most important observables in particle physics is the total cross section for hadron production in $e^{+}e^{-}$ annihilation, $\sigma(e^{+}e^{-} \rightarrow hadrons)$. It is generally presented as the ratio to the QED muon production cross section $\sigma(e^{+}e^{-} \rightarrow muons)$. The ratio is conceptually very simple for the continuum cross sections. In the quark model, 

\begin{equation}
R^{(0)}(\sqrt{s}) \equiv {\sigma(e^{+}e^{-} \rightarrow hadrons) \over \sigma(e^{+}e^{-} \rightarrow leptons)} = 3 \sum_i e^2_{q_i}
\end{equation}

with the sum extending over the quark flavors open upto the c.m. energy $\sqrt{s}$. In QCD the ratio acquires corrections in powers of $(\alpha_s/\pi)$. In the limit of massless quarks, the corrections are known to the third order, and the expansion is satisfactorily convergent.

\begin{equation}
R(\sqrt{s}) = R^{(0)}(\sqrt{s})[1 + {\alpha_s \over \pi} + C_2({\alpha_s \over \pi})^2 + C_3({\alpha_s \over \pi})^3 ...]
\end{equation}

For four open flavors ($u,d,s,c$), $C_2$ = 1.5243, and $C_3$ = -11.520 [1,2]. Because of this well established correction factor, measurements of $R(\sqrt{s})$ are often used to determine the strong coupling constant $\alpha_s$. Above the open flavor thresholds, measurements of $R$ are used to determine parameters of vector resonances. In particular, the totality of our present knowledge about charmonium resonances above the $D\bar D$ threshold at 3739 MeV comes from a single measurement of $R(\sqrt{s})$ by DASP [3]. This paper is devoted to an alternate analysis of the presently available $R(\sqrt{s})$ data for the parameters of these resonances.

The experimental determination of $R$ generally consists of measuring the total cross section for the production of hadrons, and dividing it by the lowest order QED cross section, $\sigma(e^{+}e^{-} \rightarrow muons)$ (in nb) $= 4\pi \alpha_{em}^2/3s = 86.8/s$ (GeV$^2$).

\begin{table}[tbh!]
\caption{$R$ Measurements from different sources [Ref. 3-7].}

{\small
\begin{center}
\begin{tabular}{c|c|c|c|c}
\hline
Ref. &$\sqrt{s}$ &Steps &$\%$Stat. &$\%$Syst. \\
(Year) &(GeV) &(MeV) &Error &Error \\
\hline
DASP[3] &3.6-4.8 &10-15 &$5-10$ &$\sim 12$ \\
(1978,9) & & & & \\
\hline
MARK I[4] &3.4-4.7 &10 &$5-30$ &$20-10$ \\
(1982) & & & & \\ 
\hline
PLUTO[5] &3.6-5.0 &20-40 &$3-9$ &$10-15$ \\
(1982) & & & & \\
\hline
CB[6] &3.6-4.5 &6 &$3-7$ &$\sim 9$ \\
(1986) & & & & \\
\hline
BES[7] &3.7-4.8 &5-10 &$\sim 4$ &$\sim 5$ \\
(2002) & & & & \\
\hline
\end{tabular}
\end{center}
}
\end{table}

\begin{table}[tbh!]
\caption{Parameters of the higher $\psi (1^{--})$ resonances.}

{\small
\begin{center}
\begin{tabular}{c|c|c|c}
\hline
 Source & Mass & $\Gamma_{tot}$ & $\Gamma_{ee}$ \\
 & (MeV) & (MeV) & (keV) \\
\hline
DASP[3] & $4040 \pm 10$ & $52 \pm 10$ & $0.75 \pm 0.15$ \\
(1978 & $4159 \pm 20$ & $78 \pm 20$ & $0.77 \pm 0.23$ \\
-1979) & $4417 \pm 20$ & $66 \pm 15$ & $0.49 \pm 0.13$ \\
\hline
MARK I[9] & $4417 \pm 10$ & $33 \pm 10$ & $0.44 \pm 0.14$ \\
(1976) & & & \\
\hline
PDG[1,10] & $4415 \pm 6$ & $43 \pm 15$ & $0.47 \pm 0.10$ \\
(1980-2002) & & & \\
\hline
\end{tabular}
\end{center}
}
\end{table}

The existing data for $R$, which span the region $\sqrt{s}$ = 3.8 - 4.8 GeV, which contains the resonance structures of interest, are listed in Table I along with their principal characteristics[3-7]. Radiative corrections have been made in all data sets and heavy lepton contributions have been removed. The numerical data for DASP[3] measurements have been obtained from Ref.[8].

\begin{figure}[tbh!]
\begin{center}
\vspace*{-20pt}
\includegraphics[width=9.cm,height=10cm]{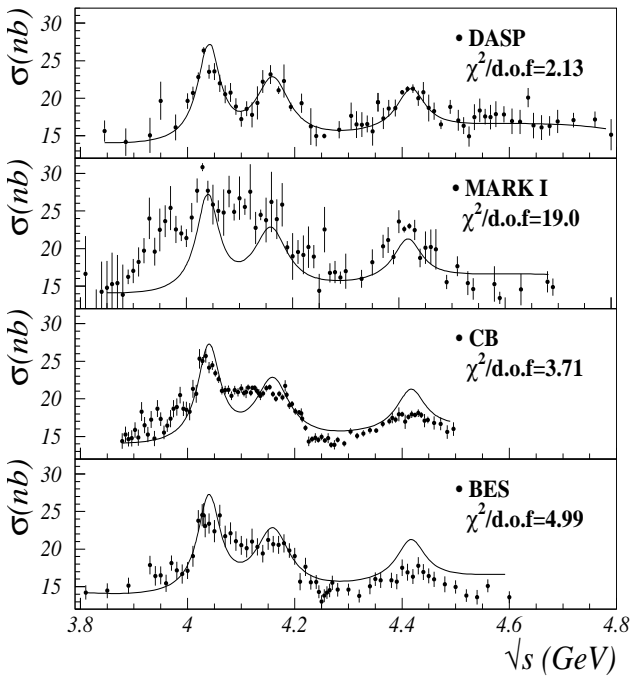}
\vspace*{-10pt}
\caption{Total hadron production cross sections in nb obtained as $\sigma(e^{+}e^{-} \rightarrow hadrons)$ = 86.8 $R/s$ (GeV) from the $R$ measurements of DASP[3], MARK I[4], Crystal Ball (CB)[6], and BES[7]. Only statistical errors are shown. The common curve superposed on all four data sets is that obtained by DASP[3] as the best fit to their data.}
\end{center}
\vspace*{-20pt}
\end{figure}

Because we wish to determine Breit-Wigner resonance parameters we have converted the $R$ values to $\sigma(e^{+}e^{-})$ in nb = 86.8 $R$/$s$ (in GeV). These cross sections are plotted in the four panels of Fig. 1. We do not use the PLUTO[5] data because the rather coarse spacing of its data points, 20--40 MeV, does not allow analysis for resonances whose widths are of the same order.  Only DASP [3] has reported a resonance analysis of their data, and in Fig. 1 we have superposed on all data sets the curve which represents the fit obtained by DASP[3] by fitting their data with three Breit-Wigner resonances and a background smoothly connecting their measured $R$ values at the two extremes of 3.6 GeV and 5.0 GeV. The $\chi^2/d.o.f.$ values listed in the panels of Fig. 1 indicate that the fits with DASP parameters are quite poor in all other cases, particularly to the Mark I data.  The resonance parameters obtained by DASP are listed in Table II. The Particle Data Group (PDG) adopted the DASP parameters for the resonances at 4.04 and 4.16 GeV, and averaged the DASP parameters for the third resonance with those given in an early report by MARK I[9]. These are the only available resonance parameters, and they have remained unchanged from PDG-1980[10] to PDG-2004[1].

\begin{figure}[tb!]
\begin{center}
\vspace*{-18pt}
\includegraphics[width=9.cm]{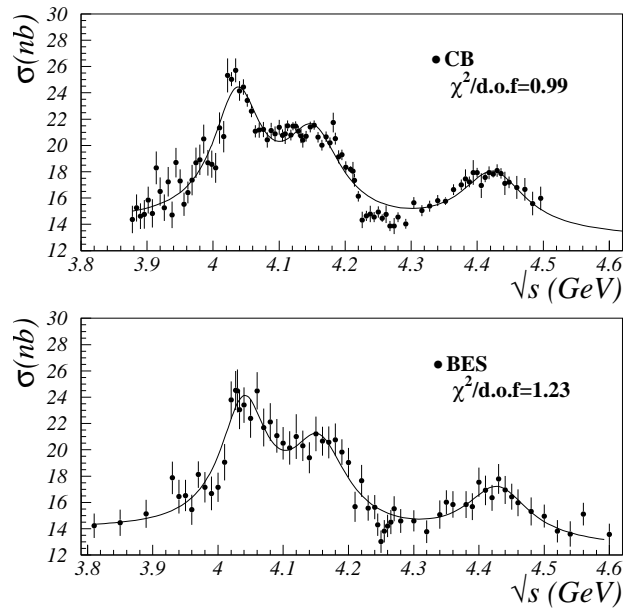}
\vspace*{-10pt}
\caption{Total hadron production cross sections and the $\chi^2/d.o.f$ for the best fits shown: (Top) Crystal Ball (CB)[6], (Bottom) BES[7].}
\end{center}
\vspace*{-20pt}
\end{figure}

 The three pre-1990 data sets shown in Fig. 1 differ both qualitatively and quantitatively. This was noted early by, for example, Eichten {\it et al.}[11], and seems to have discouraged theoretical attempts to understand higher vector states of charmonium. One of the most obvious differences is that neither MARK I, nor CB, nor BES find the deep minimum at $\sim$ 4.1 GeV observed in the DASP data[12]. Similarly, all three appear to observe broader structures than those claimed by DASP, and in the initial report by MARK I[9]. There are also obvious differences in the absolute levels of cross-sections which can be best compared in the off-resonance regions.  In the region $\sqrt{s}=3.60-3.65$, Mark I[4] cross sections are $\sim27\%$ larger than those of DASP [3].  Similarly, in the off-resonance region, $6.0-7.5$ GeV, where additional data from CB are available [13], Mark I data [4] are $\sim28\%$ larger than CB.  The only two data sets which are not only in qualitative, but excellent quantitative agreement, are those due to CB [6] and BES [7]. They also have the smallest step size and the smallest statistical and systematic errors.  We base the present resonance parameter determination on these two data sets.

In order to obtain the best resonance parameters possible, we fit the separate CB and BES data sets, each with three Breit-Wigner resonances, with $$\sigma_{BW}(\sqrt{s})=(3\pi/4p^2)\Gamma_{ee}\Gamma_h/[(\sqrt{s}-M)^2+(\Gamma_{tot}/2)^2],$$ where $\Gamma_{ee}$, $\Gamma_h$, and $\Gamma_{tot}$ are the mass independent electron, hadron and total widths, respectively, for a vector resonance of mass $M$ produced in the head-on collision of $e^+$ and $e^-$, each of momentum $p$.  There is little evidence in the data for any substantial variation of the continuum background in this region of $\sqrt{s}$. We therefore parametrize it with a linear function, $\sigma(bkgd) = A - B(\sqrt{s} - 3.8$ GeV). The results of the separate best fits are listed in Table III and the fits are shown in Fig. 2. As expected, the possible normalization differences between CB and BES, and their individual absolute cross section uncertainties, quoted as systematic errors in Table 1, do not reflect in the masses or the total widths; they are found to be in near perfect agreement. Also, the differences in electron widths are well within errors. It is very gratifying that two independent measurements, made twenty years apart, are in such good agreement. Because of this excellent agreement in the results of the CB and BES measurements, we consider their weighted averages, given in the last column of Table III, as our final results.

\begin{table}[tb!]
\caption{Summary of results. Masses $M^{(i)}$ and total widths $\Gamma_{tot}^{(i)}$ are in MeV, electron widths $\Gamma_{ee}^{(i)}$ are in keV, the parameter $A$ is in nb, and the parameter $B$ is in nb/GeV. The final row shows the $\chi^2$/d.o.f. for each fit.}

{\small
\begin{center}
\begin{tabular}{|c|c|c|c|}
\hline
\hline
 & $M^{(1)}$ & $\Gamma^{(1)}_{tot}$ & $\Gamma^{(1)}_{ee}$ \\
 & (MeV) & (MeV) & (keV) \\
\hline
PDG[1] & $4040 \pm 10$ & $52 \pm 10$ & $0.75 \pm 0.15$ \\
CB[6] & $4037 \pm 2$ & $85 \pm 10$ & $0.88 \pm 0.11$ \\
BES[7] & $4040 \pm 1$ & $89 \pm 6$ & $0.91 \pm 0.13$ \\
\hline
CB+BES & $4039.4 \pm 0.9$ & $88 \pm 5$ & $0.89 \pm 0.08$ \\
\hline
\hline
 & $M^{(2)}$ & $\Gamma^{(2)}_{tot}$ & $\Gamma^{(2)}_{ee}$ \\
\hline
PDG[1] & $4159 \pm 20$ & $78 \pm 20$ & $0.77 \pm 0.23$ \\
CB[6] & $4151 \pm 4$ & $107 \pm 10$ & $0.83 \pm 0.08$ \\
BES[7] & $4155 \pm 5$ & $107 \pm 16$ & $0.84 \pm 0.13$ \\
\hline
CB+BES & $4153 \pm 3$ & $107 \pm 8$ &  $0.83 \pm 0.07$\\
\hline
\hline
 & $M^{(3)}$ & $\Gamma^{(3)}_{tot}$ & $\Gamma^{(3)}_{ee}$ \\
\hline
PDG[1] & $4415 \pm 6$ & $43 \pm 15$ & $0.47 \pm 0.10$ \\
CB[6] & $4425 \pm 6$ & $119 \pm 16$ & $0.72 \pm 0.11$ \\
BES[7] & $4429 \pm 9$ & $118 \pm 35$ & $0.64 \pm 0.23$ \\
\hline
CB+BES & $4426 \pm 5$ & $119 \pm 15$ &  $0.71 \pm 0.10$\\
\hline
\hline
 & $A$ & $B$ & $\chi^2$ \\
\hline
DASP[3] & Polynomial & & 2.1 \\
CB[6]& $14.2 \pm 3.5$ &  $1.5 \pm 0.4$ & 0.99 \\
BES[7] & $13.7 \pm 4.5$ & $1.5 \pm 0.5$ & 1.23 \\
\hline
\hline
\end{tabular}
\end{center}
}
\end{table}

From Table III it is apparent that the presently determined masses of the three resonances have much smaller errors than the masses determined by DASP[3], but are in general agreement with them. However, the total widths determined in the present analyses are quite different from those determined by DASP, and those adopted by PDG2004. Our values of $\Gamma^{(1)}$, $\Gamma^{(2)}$ and $\Gamma^{(3)}$ and $\sim 67\%$, $37\%$, and $179\%$ larger, respectively than the corresponding PDG values. The corresponding electron widths determined by us are also larger, by $23\%$, $8\%$, and $51\%$. The errors in our width determinations are also more than factor two smaller. It is also worth noting that the present analysis does not rule out the presence of narrow structures other than the three broad ones analyzed here.

It is interesting to note that the average value of $R$ over the entire region of measurements is $<R> = 3.63 \pm 0.13$ for CB, and $<R> = 3.35 \pm 0.14$ for BES. These values are close to the prediction, $R = 3.67$ obtained from eq. 2 for four flavors, with $\alpha_s = 0.30$, which corresponds to $\alpha_s(m_Z) = 0.117$[1].

Eichten {\it et al.} have identified, and most later potential model calculations agree with their identification, that the three resonances are $3 ^3S_1$ (4039 MeV), $2 ^3D_1$ (4153 MeV), and $4 ^3S_1$ (4426 MeV). Three early potential model calculations have made predictions of the leptonic widths of $3 ^3S_1$ and $4 ^3S_1$ states. Quigg and Rosner predicted $\Gamma_{ee}$(4039) = 1.00 keV, and $\Gamma_{ee}$(4426) = 0.51 keV for their logarithmic potential[14]. Eichten {\it et al.}[11] predict $\Gamma_{ee}$(4039) = 1.5 keV, and $\Gamma_{ee}$(4426) = 1.1 keV with the Cornell potential, and Buchm\"uller and Tye[15] predict $\Gamma_{ee}$(4039) = 1.68 keV, and $\Gamma_{ee}$(4426) = 1.31 keV (assuming $\Gamma_{ee}(J/\psi)$ = 5.25 keV) with a QCD based potential. In principle, all other potential model calculations give predictions for wave functions and therefore leptonic widths, but no subsequent explicit predictions can be found in the published literature.

Theoretical studies of the hadronic properties of the region above the $D\bar D$ threshold are extremely rare, mainly due to the fact that there were ``obvious disagreements between experiments''[11]. Despite this problem, Eichten {\it et al.}[11] made a rather detailed, though unavoidably conjectural, study of this region. They attempted to explain the nature of the observed structures in the 3.9 to 4.2 GeV region in terms of the 3$S$ and 2$D$ resonances, the mixing between them, and the opening of the $\bar DD^{*}$ and $D^{*}\bar D^{*}$ thresholds at 3880 MeV and 4020 MeV respectively. They did not succeed in reproducing the deep minimum observed in the data of DASP, but as we have seen, the CB and BES data definitely rule out the minimum, and our resonance parameters fit these data very well. Eichten {\it et al.} explained the structure at 4.4 GeV entirely in terms of the 4$S$ resonance and its $D^{*}\bar D^{*}$ decay. The discovery of strange $D_s$ and $D^{*}_s$ came later, and the opening of the $D_s\bar D_s$ and $D^{*}_s\bar D^{*}_s$ thresholds at 3940 MeV and 4224 MeV, respectively, can be certainly expected to play a significant role in the understanding of the observed structures, particularly that of the 4$S$ state at 4426 MeV. We hope that our clarification of the experiment data, and the more precise determination of the resonance parameters will lead to renewed theoretical interest in the understanding of the challenging physics of these higher vector states of charmonium.
  

This work was supported by the U.S. Department of Energy. The author wishes to thank Dr. Ismail Uman and David Joffe for their help in the preparation of this paper. 



\begin{thebibliography}{14}
\bibitem{12} Both MARK I and CB authors suggest that the deep minimum seen in the DASP data results from biased radiative corrections.
\end{thebibliography}
\end{document}